\documentclass[twocolumn,10pt,aps,pra,showpacs,superscriptaddress,nobalancelastpage,longbibliography,nofootinbib,floatfix]{revtex4-2}

\usepackage[pdftex]{graphicx}
\usepackage{bm}
\usepackage{color}
\usepackage{MnSymbol}
\usepackage{enumerate}
\usepackage{bbold,soul}
\usepackage{lipsum}
\usepackage{array}
\usepackage{hyperref}
\usepackage{textgreek}
\usepackage[dvipsnames]{xcolor}
\usepackage{tabularx}
\usepackage{graphics,epsfig}
\usepackage{wrapfig}
\usepackage{mathrsfs}
\usepackage{cleveref}
\usepackage{tikz}
\usepackage{algorithm}
\usepackage{algpseudocode}
\usepackage{booktabs}

\usetikzlibrary{quantikz}
\usepackage{xcolor}
\usepackage{pgfplots}
\pgfplotsset{compat=1.18}

\definecolor{qubit}{HTML}{4169E1}
\definecolor{algo}{HTML}{228B22}
\definecolor{attack}{HTML}{B31B1B}

\begin{document}
\title{Crosstalk Attacks and Defence in a Shared Quantum Computing Environment}

\author{Benjamin Harper}
\affiliation{School of Physics, The University of Melbourne, Parkville, 3010, Victoria Australia}
\affiliation{Quantum Systems, Data61, CSIRO, Clayton, 3168, Victoria Australia}

\author{Behnam Tonekaboni}
\affiliation{Quantum Systems, Data61, CSIRO, Clayton, 3168, Victoria Australia}

\author{Bahar Goldozian}
\affiliation{Quantum Systems, Data61, CSIRO, Clayton, 3168, Victoria Australia}

\author{Martin Sevior}
\affiliation{School of Physics, The University of Melbourne, Parkville, 3010, Victoria Australia}

\author{Muhammad Usman}
\affiliation{School of Physics, The University of Melbourne, Parkville, 3010, Victoria Australia}
\affiliation{Quantum Systems, Data61, CSIRO, Clayton, 3168, Victoria Australia}

\begin{abstract}
Quantum computing has the potential to provide solutions to problems that are intractable on classical computers, but the accuracy of the current generation of quantum computers suffer from the impact of noise or errors such as leakage, crosstalk, dephasing, and amplitude damping among others. As the access to quantum computers is almost exclusively in a shared environment through cloud-based services, it is possible that an adversary can exploit crosstalk noise to disrupt quantum computations on nearby qubits, even carefully designing quantum circuits to purposely lead to wrong answers. In this paper, we analyze the extent and characteristics of crosstalk noise through tomography conducted on IBM Quantum computers, leading to an enhanced crosstalk simulation model. Our results indicate that crosstalk noise is a significant source of errors on IBM quantum hardware, making crosstalk based attack a viable threat to quantum computing in a shared environment. Based on our crosstalk simulator benchmarked against IBM hardware, we assess the impact of crosstalk attacks and develop strategies for mitigating crosstalk effects. Through a systematic set of simulations, we assess the effectiveness of three crosstalk attack mitigation strategies, namely circuit separation, qubit allocation optimization via reinforcement learning, and the use of spectator qubits, and show that they all overcome crosstalk attack risk with varying degrees of success and help to secure quantum computing in a shared platform.
\end{abstract}

\maketitle

\section{Introduction}
In the pursuit of achieving scalability and performance in quantum computing, efficient control and mitigation of noise are essential for the high-fidelity implementation of quantum algorithms of practical interest. Noise in quantum computers may take several forms including environmental noise that causes decoherence in qubits via dephasing and relaxation processes; also, computational gates and measurement operations on qubits may result in imperfect fidelities \cite{AlvarezSuterReview, ClerkReview}. In addition to the aforementioned sources of noise, qubits connectivity---which is essential in executing multi-qubit gates---could lead to the emergence of unwanted correlated dynamics. These correlated dynamics give rise to an important type of noise known as crosstalk noise \cite{Sarovar2020, White2023, Gambetta2018, Sung2021},  which results in a decline in quantum coherence and the introduction of errors in quantum computations.

\begin{figure*}[ht]
 \includegraphics[width=0.95\linewidth]{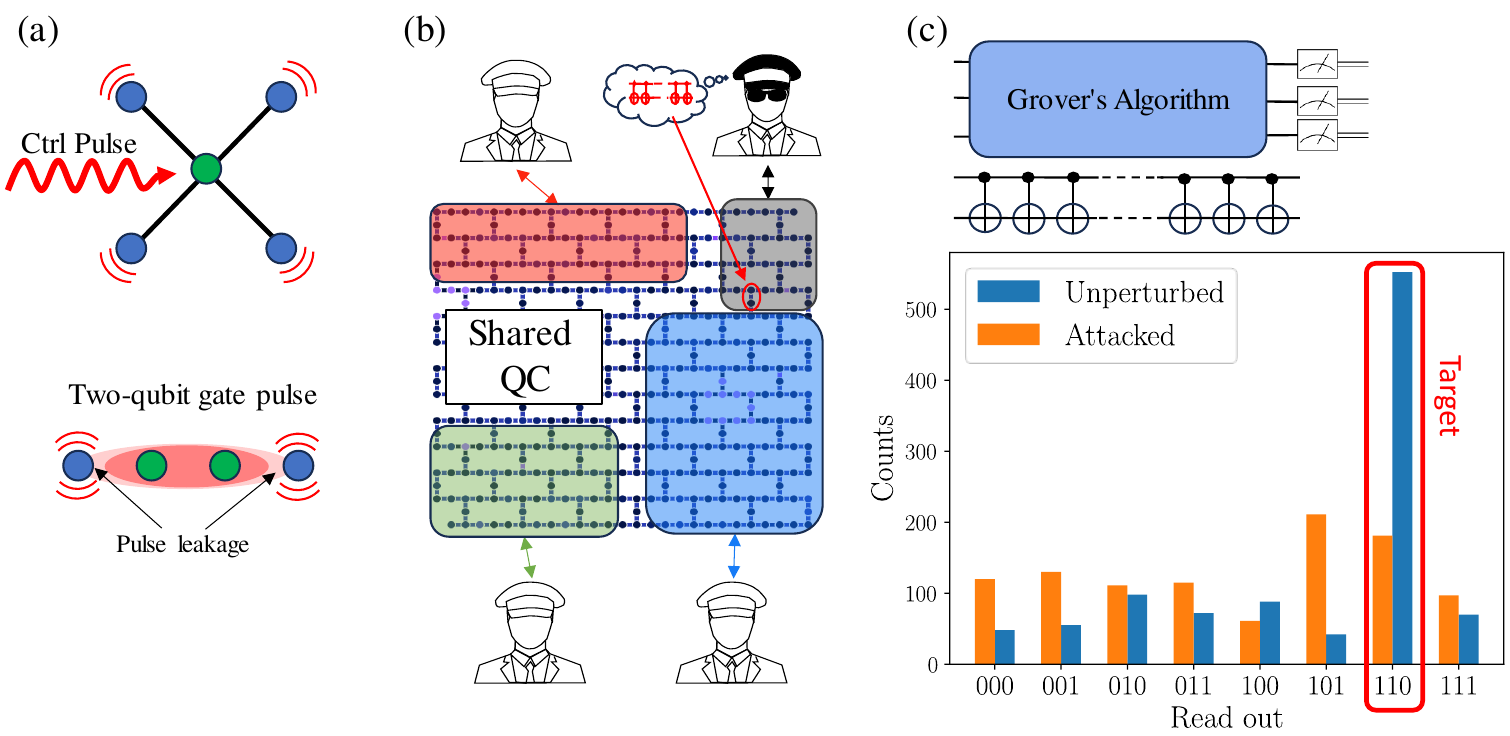}
    \caption{(a) Illustration of crosstalk effect mechanism. (Top) shows the crosstalk effect due to the always on coupling between qubits, the coupling is demonstrated with solid lines. Although the control pulse (red wavy arrow) aims only the target qubit (green) in the middle, neighbour qubits (blue) are affected. (Bottom) depicts the leakage of a two-qubit gate that aims only two middle qubits (green). Although there is no link between the qubits, the leakage of the two-qubit gate affects the neighbour qubits.
    (b) Illustration of a future shared Quantum Computer with multiple users each running their own programs. A malicious actor has gained access and can employ crosstalk attacks against other programs (background is IBM 433-Qubit Osprey device). (c) Top: We implement Grover’s algorithm with the target of ``$6$'' on a real 5-qubit IBM Quantum computer. Qubits $1$-$3$ implement Grover’s algorithm while on qubits $4$ and $5$ we implement a series of CNOT gates. (Bottom) When the qubits $3$ and $4$ are not operating, the circuit returns the blue histogram, showing normal operation. When we also apply CNOT operations on qubits $3$ and $4$, we instead observe the orange histogram showing crosstalk from the CNOT gates completely overwhelming the operation of the circuit.}
    \label{fig:Intro}
\end{figure*}
It is noted that the word \textit{crosstalk} is often used in the fields of electrical engineering and communication theory to describe the phenomenon of unwanted signal coupling across different pathways \cite{mazda2014telecommunications}. However, in quantum computing, the term is adapted to describe a variety of physical phenomena when an undesired influence is exerted by one subsystem of an experimental apparatus, such as a qubit or control pulse, on another qubit. Consider superconducting qubits, such as those used in IBM devices, as an example. The qubits in these devices are placed in a certain topology, with each qubit connected to one, two, or more neighbor qubits. A cross-resonance pulse will then be utilized to apply a two-qubit gate to a pair of linked qubits~\cite{Gambetta2018, Sung2021}. While this \textit{always-on} coupling, utilises the two-qubit gates, they also result in an undesirable crosstalk effect. This is depicted in the top image of FIG.~\ref{fig:Intro}(a), where a single qubit gate pulse targets only the center qubit, but due to physical coupling, there is undesired crosstalk on the neighbor qubits. Even with quantum platforms without always-on coupling, such as atomic qubits, the profile of the laser gate pulse may leak to the side atoms and adversely affect them~\cite{Fang2022}. The bottom image of FIG.~\ref{fig:Intro}(a) illustrates this mechanism.   

While crosstalk is expected to be present in all quantum devices and suppressing crosstalk in quantum computers is an active topic of research (see~\cite{Zhou2023,parrado2021crosstalk,Buterakos2018} as examples), here we investigate crosstalk as a substantial security concern. It is almost certain that functional quantum computers will be operated as multi-user facilities, executing circuits for multiple users simultaneously, including public access through cloud platforms. Since the operations of neighboring circuits affect each other through crosstalk, it is possible to craft an algorithm that can interfere with nearby normal operations, causing programs to fail, provide faulty results, or leak information. This could be exploited by an adversary or deployed against an adversary’s quantum computer operations \cite{saki_survey_2021,deshpande2022antivirus,gaur2024crosstalkattackresilientrns}. An illustration of one such scenario is shown in FIG.~\ref{fig:Intro}(b) exhibiting a malicious actor interfering with an algorithm employed on a multi-user quantum computer. It may also be possible for an adversary to extract private information through crosstalk, in analogy to a classical crosstalk attack~\cite{9676573}.

To demonstrate a crosstalk attack, we executed a three-qubit Grover algorithm with a target answer of `$110$', on an IBM device. The demonstration encompassed two scenarios: one without any external interference and the other with the imposition of a crosstalk attack. The crosstalk attack was implemented through a series of CNOT gates on two proximate qubits. Top row of FIG.~\ref{fig:Intro}(c) shows the circuit diagram of the demonstration and the histogram in this figure shows the results of the demonstration. In the absence of any attack, the majority of counts are the target answer. In contrast, when subjected to the attack, the target result is wiped out and the histogram shows an approximately uniform distribution of counts.

It is important to note that crosstalk attacks are distinctly different from malware-based attacks on conventional computers. Crosstalk noise is inherent in the actual physics of the operation of the quantum computer and this kind of attack vector is simply not possible in conventional computers. Therefore, defending against these attacks requires strategies in addition to normal security procedures. 

Despite significant evidence highlighting the potential security implications of crosstalk attacks on quantum processors~\cite{saki_survey_2021}, there is currently a lack of quantitative modelling and practical mitigation strategies for real devices. Addressing this gap is the primary objective of this paper. The structure of the paper is as follows. First in Sec.~\ref{sec:characterization}, we characterize crosstalk noise on an IBM device. This characterization will be the framework for simulating circuits under the effects of crosstalk noise in the rest of the paper. Then, in Sec.~\ref{sec:mitigation}, we present three techniques to detect and mitigate the effects of a crosstalk attack on a circuit. The first technique (Sec.~\ref{sec:mitigation}A) employs unused buffer qubits to distance the circuits from potential crosstalk attacks. This technique also provides an indication of the extent of crosstalk noise in an IBM device, laying the foundation for basic understanding of the severity of such noise mechanism. The second technique (Sec.~\ref{sec:mitigation}B) employs a reinforcement machine learning approach and uses the noise model of the device to position circuits such that the  potential for crosstalk noise to impact their performance is minimised. The final technique (Sec.~\ref{sec:mitigation}C) employs extra qubits, but in contrast to first technique these extra qubits are used as spectator qubits. By measuring the spectator qubits at the correct interval, it is possible to detect the presence of an attack with a high level of confidence. Finally, in Sec.~\ref{sec:conclusion}, we discuss the importance of the security implications of crosstalk attacks and our successful strategies to mitigate them.

\section{Characterising and Simulation of Crosstalk on IBM Devices}\label{sec:characterization}
Essential to our work in this paper is a realistic characterisation of the noise due to crosstalk in a physical device. For this, we opted to use idle tomography~\cite{blume-kohout_idle_2019}. The Python library PyGSTi~\cite{erik_pygsti_2016} is used to perform the idle tomography calculations, and also simulate circuits with crosstalk noise. We benchmark crosstalk on a real quantum device, ibm\_hanoi, and use this data to prepare the crosstalk simulator. We then run validation circuits on the crosstalk simulator, the real device, and other existing quantum software simulators to verify that the crosstalk simulator is an accurate representation of real devices.

\begin{figure*}
    \centering
    \begin{tikzpicture}
        \node at (-5, 1.8) {(a)};
        \node at (-1, 1.5) {Idle Tomography Circuit};
        \node[scale=0.85] at (-1, 0) {
            \begin{quantikz}
                & \qwbundle{n-2} & \gate{\text{Pauli}} & \qw{} & \qw{} & \qw{} & \qw{} & \qw{} &
                \gate{\text{Pauli}_2} &\meter{} \\
                & \lstick{$q_a$} & \qw{} & \ctrl{1} & \qw{} & \ldots\ & & \ctrl{1} & \qw{} & \qw{} \\
                & \lstick{$q_b$} & \qw{} & \targ{} & \qw{} & \ldots\ & & \targ{} & \qw{} & \qw{} \\
            \end{quantikz}
        };

        \draw[rounded corners, dashed] (-2.7, -0.95) rectangle (0.3, 1.1) {};
        \node at (-1.2, -1.15) {Idle Period};
        
        \node at (-0.8, -4.2) {\includegraphics[width=0.45\textwidth]{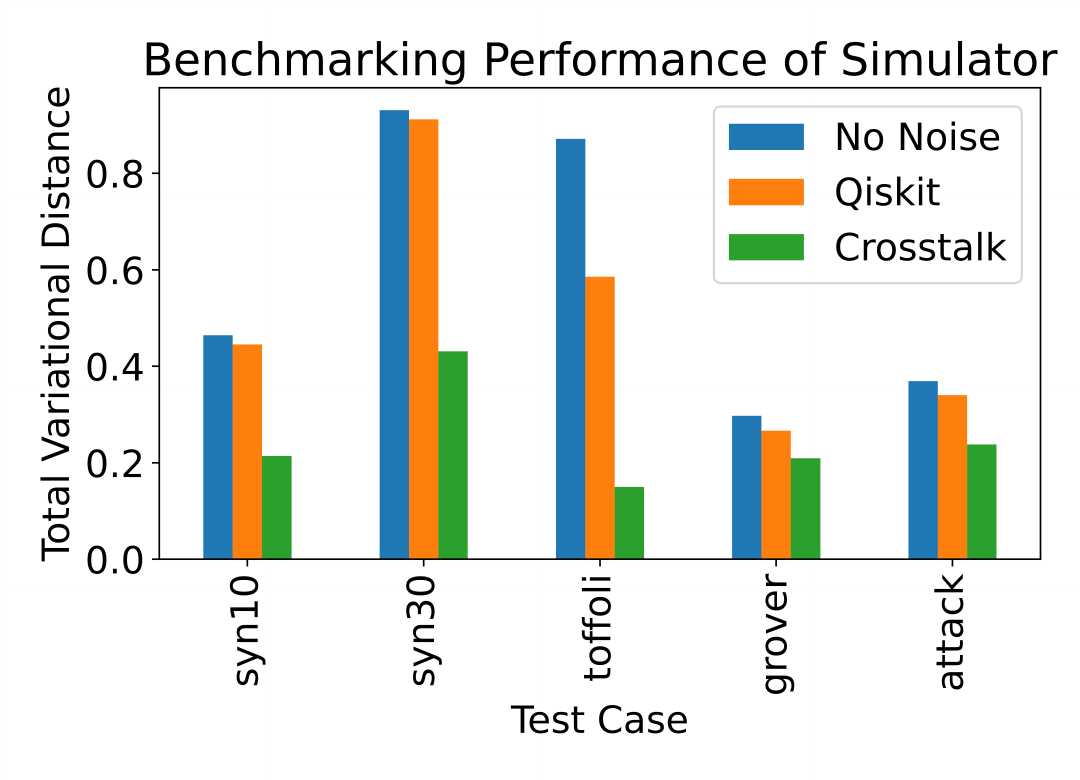}};
        \node at (-5, -1.2) {(b)};

        % toffoli
        \node at (8, -0.25) {\includegraphics[width=0.45\textwidth]{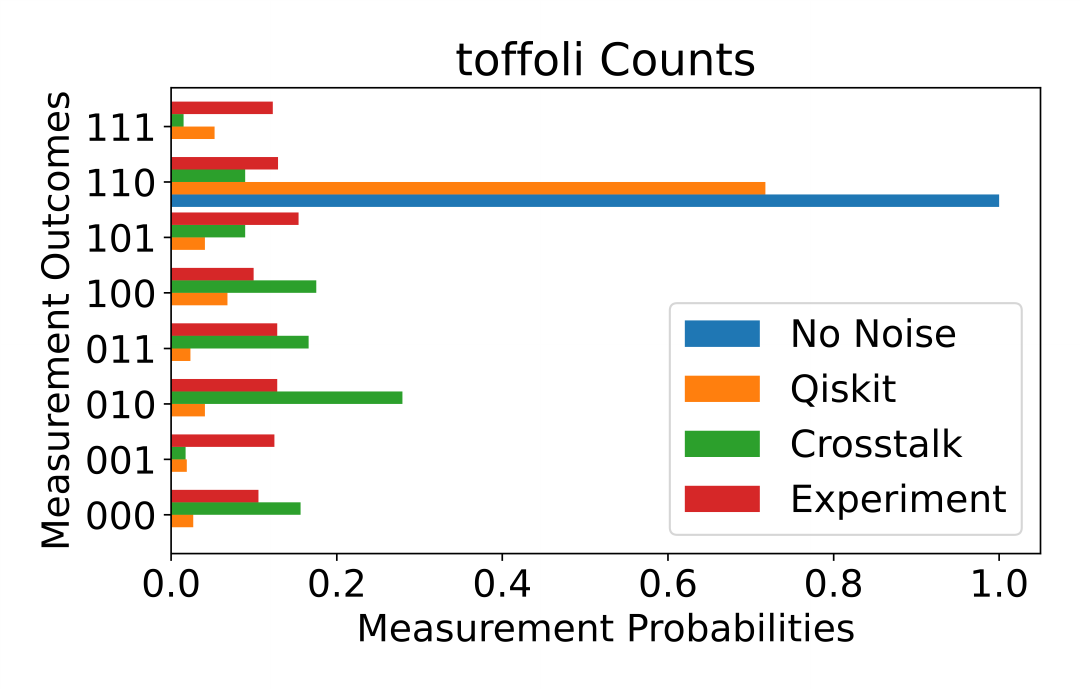}};
        \node at (3.8, 1.8) {(c)};
        
        % grover
        \node at (8, -5.1) {\includegraphics[width=0.45\textwidth]{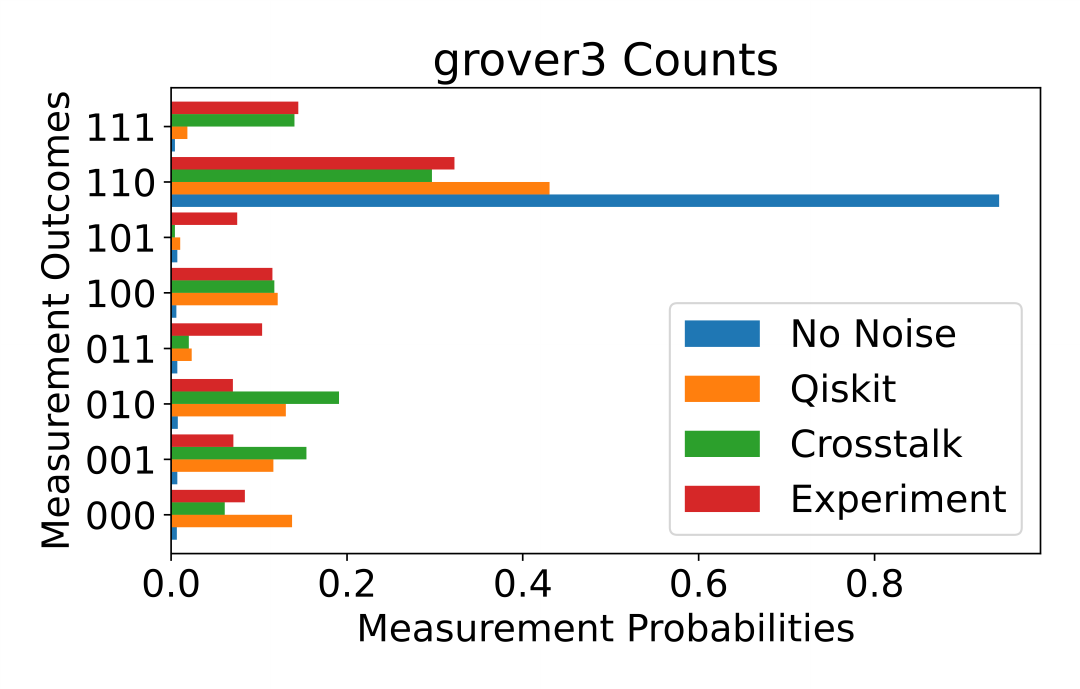}};
        \node at (3.8, -2.8) {(d)};
    \end{tikzpicture}
        \caption{(a)~Example of an idle tomography circuit for benchmarking crosstalk. Here we are benchmarking the crosstalk due to a CNOT between qubits $q_a$ and $q_b$, with idle tomography being performed on the remaining $n-2$ qubits on the device. The tomography qubits are prepared into a Pauli basis state, then left to idle while CNOTs are applied between qubits $q_a$ and $q_b$. At the end of the circuit, the state is measured in a Pauli basis which may be distinct from the originally prepared Pauli basis. (b) We now investigate the performance of our simulator through benchmarking. A variety of benchmark circuits are run on a real device, and three simulators. A description of the simulation techniques and benchmark circuits is provided in the main text. Performance of the simulation techniques is judged using Total Variational Distance (see text), with the device noise set as the baseline. Lower is better (closer to real device). (c) and (d) present the measured counts in two of the test cases, (c) Toffoli, (d) Grover. In both cases the target circuit output is $\ket{110}$. The target output distribution for each simulation technique is to match the demonstration results.}\label{fig:idt_and_sim}
\end{figure*}
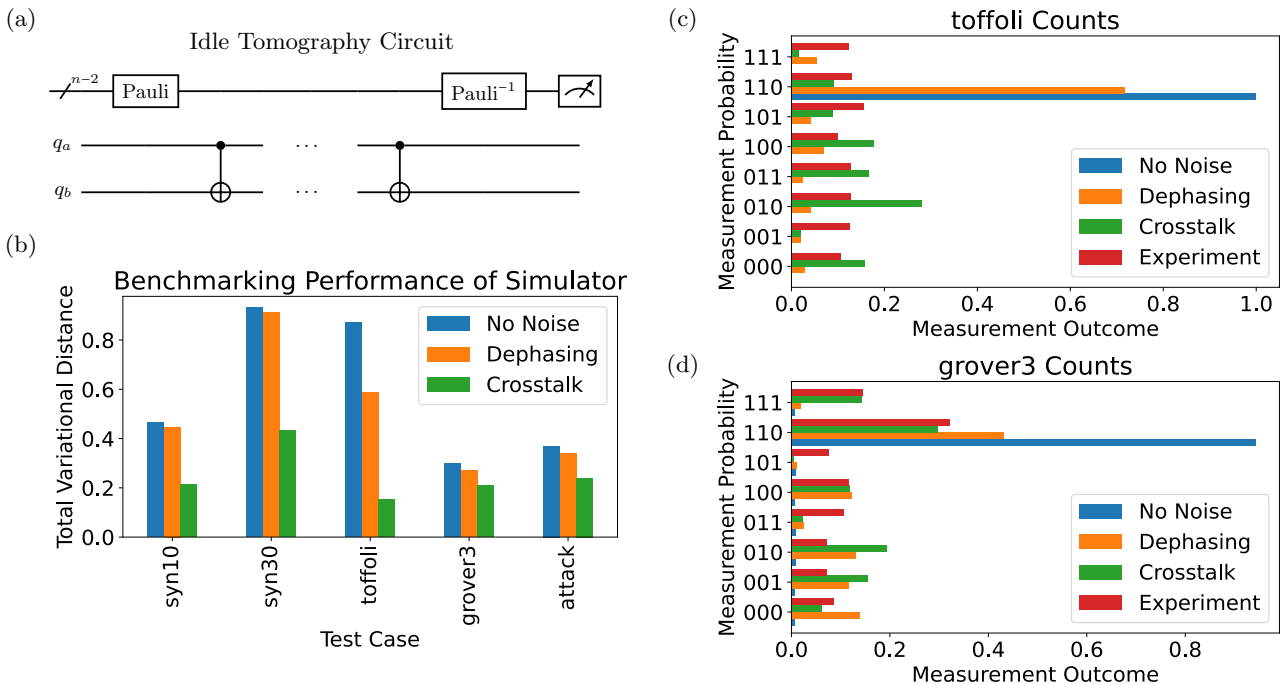

\subsection{Idle Tomography}
Idle tomography (IDT)~\cite{blume-kohout_idle_2019}, as the name suggests, seeks to characterise the idle error rates of qubits on a physical device. To do this, a family of tomography circuits is prepared. The circuits we used consist of three parts; preparation, idling and un-preparation, as shown in FIG.~\ref{fig:idt_and_sim}(a). In the preparation step, each circuit initialised in the $\ket{0}$ state and is prepared in a unique Pauli basis. In the second phase, the qubits are left to idle for a variable period of time (measured in gate lengths). Finally, the state is measured in a Pauli basis, which may be different to the originally prepared basis. In a noise-free quantum computer, this returns the state to $\ket{0}^n$. In a noisy device, however, the measurement results will be perturbed away from the target distribution. A large number of circuits in this form are generated, where only the Pauli state preparation is changed.

The noise channel modelled by idle tomography in PyGSTi is the reduced HSA model~\cite{blume-kohout_idle_2019},
\begin{equation}
    \mathcal{L} = \mathcal{H} + \mathcal{S} + \mathcal{A}.
\end{equation}

This model has three components; Hamiltonian errors ($\mathcal{H}$) are the rate of over (under) rotation induced on a specific Pauli axis. The stochastic error rate ($\mathcal{S}$) is the probability of a random Pauli $X$, $Y$ or $Z$ error. Finally affine errors ($\mathcal{A}$) correspond to non-unitary noise processes.

The information contained in the probability distribution of a circuit's output scales exponentially with the number of qubits --- as there are an exponential number of outcomes. Conversely, the number of parameters in the reduced HSA model scales polynomially with the number of qubits in the device. As such, the number of experiments which need to be performed to benchmark a device increases logarithmically with the number of qubits~\cite{blume-kohout_idle_2019}, though the number of shots for each experiment to obtain a good probability distribution increases exponentially. While randomised benchmarking~\cite{Emerson_2005} is more sample efficient than idle tomography and often used for device benchmarking, it smooths noise channels into an average error rate, and cannot capture the subtle dynamics of crosstalk noise~\cite{ash-_saki_experimental_2020}. The extra overhead of IDT is required for applications that consider crosstalk noise specifically. Regarding the scalability of the IDT method, we note that only small sized patches of quantum devices are required to run NISQ algorithms which can be charaterised and bechmarking using IDT and our crosstalk mitigation techniques.

When considering the noise due to gates on a quantum computer, we typically look at the qubits on which the gate is applied. In crosstalk noise, we are interested in the noise that a gate causes on \textit{other} qubits around it. In effect, we want to know the idle error rate of qubits in the device when a gate is being applied to neighbouring qubits. Thus, to quantify the crosstalk error rates due to an $m$-qubit gate $U$ on an $n$-qubit device, we choose $m$ qubits on which we want to benchmark the gate. On the remaining $n-m$ qubits, idle tomography is performed. While the IDT qubits are idling, the gate $U$ is performed in parallel. The presence of crosstalk in the device means that $U$ will cause errors in the IDT qubits.

In this work, we benchmark a 27-qubit IBM device --- ibm\_hanoi. We benchmark crosstalk due to the 2-qubit CNOT gate. Hence, $m=2, n=27-2=25$. This requires 765 IDT circuits to be run. These results were then postprocessed classically to produce a realistic simulator with crosstalk noise, using PyGSTi's \texttt{cloud\_crosstalk\_model} simulator. The IBM calibration parameters for the device at the time of benchmarking are shown in appendix~\ref{app:device-params}.

\subsection{Simulation of Crosstalk}\label{sec:benchmarking}
Having created a crosstalk-aware classical simulator, it is necessary that we validate that its performance matches that of the real device it is based on. To perform this validation of the crosstalk simulator, we ran five benchmark circuits and compared the results to the real quantum device, as well as other simulation methods. Ideally, our simulator will be a better model of crosstalk effects in the real devices compared to other simulation noise models which represent generic noise such as depolarisation or other effects, especially in circuits with a large amount of crosstalk, induced by CNOT gates. The output of these experiments is a set of counts, which is used to approximate the probability distribution of that quantum circuit. A good simulator should have a similar probability distribution to the output distribution of the real device. The metric used to measure performance is the Total Variational Distance (TVD)~\cite{ash-_saki_experimental_2020}, which measures the distance between probability distributions in the $L_\infty$ metric. The TVD is explained further in appendix~\ref{app:tvd}.

In FIG.~\ref{fig:idt_and_sim}(b) we show TVD relative to the real device for three scenarios: 1) \textit{No Noise}, a noise free simulation of the circuit, 2) \textit{Qiskit Noise}, simulation using the noise from the builtin noise model in IBM's Qiskit package which does not include crosstalk, but does include gate errors, T1-relaxation and T2-dephasing times, and 3) \textit{Crosstalk Noise}, the noisy simulation that we developed for this paper. The TVD is measured relative to the results from running each circuit on the real device. As such, a lower TVD means that the simulator is a better representation of the real device. In this case, crosstalk noise is a significant source of noise in the device, hence the crosstalk simulator is closest in TVD to the real device.
We then compare these three scenarios on five different test cases: 1) The \textit{syn10} and 2) \textit{syn30} test cases are synthetic circuits of a repeated CNOT next to an idling qubit --- similar to the idle tomography circuits used to derive the noise models, 3) the \textit{grover3} test is a 3-qubit Grover's algorithm with two iterations, 4) the \textit{toffoli} test is a sequence of 10 repeated Toffoli gates, which is itself decomposed into CNOT gates and single qubit rotations, 5) the \textit{attack} test case is a Grover's algorithm circuit in the presence of a CNOT attack. Circuit diagrams for these test circuits are given in appendix~\ref{app:circ}. While the \textit{syn10. syn30} and \textit{attack} circuits are closely related to the IDT circuits and the crosstalk attack applications considered in this work, the Toffoli decomposition and Grover's algorithm circuits are demonstrations of general circuits, showing the applicability of our simulation techniques to general noisy circuit simulation. FIG.~\ref{fig:idt_and_sim}(c) and (d) show the simulated probability distribution for two of these test cases, illustrating how the crosstalk simulator counts are much closer to the real device than other simulation techniques.

\section{Crosstalk Detection and Mitigation}\label{sec:mitigation}
In this section we explore various techniques to detect and/or mitigate a crosstalk attack. For all the techniques discussed, we use the same model of attack; that is, a hostile user with the ability to craft a repeated CNOT circuit (illustrated in FIG.~\ref{fig:Intro}(c)) maximising crosstalk. We note that in principle a variety of attacks can be designed by using different types of quantum gates or their combinations, as many quantum gates can induce crosstalk~\cite{ohkura_simultaneous_2022}. However, in this work we use only CNOT gates as a proof-of-concept demonstration.

\subsection{Mitigation by Circuit Separation}\label{separation}
We begin with a study of how the strength of crosstalk changes across the device. While generally crosstalk is a non-local effect, such that quantum gates may affect neighbouring qubits (contrasted with a local effect, where a gate only induces noise to the qubits on which it was applied), we expect it to be relatively local, i.e. the qubits most affected by crosstalk from an operation are the ones physically closest to that operation. Here we define the radius of separation, $r$, to be the number of idle qubits on the shortest path between two algorithms on a device's graph of connectivity. Note that this may not correspond perfectly to the physical distance between qubits, it is a useful approximation. A recent study~\cite{ohkura_simultaneous_2022} has found that while adjacent computations can adversely affect each other, only a small radius of separation is required to mitigate this. Other work~\cite{deshpande2022antivirus,inproceedings} has found that a one qubit buffer around circuits is a useful defence against crosstalk attacks. Our work extends and combines these two results, by exploring how the effectiveness of a crosstalk attack drops as the distance between the attack and victim circuits is increased.

To test this, we created a circuit containing Grover's algorithm, and placed next to it a repeated CNOT attack as described above, where every layer in the Grover circut is accompanied by a CNOT gate on the neighbouring attack qubits. The circuit was placed at the centre as in figure~\ref{fig:separation}(a), on qubits 12, 13, and 14. The attack was then placed at all possible locations around the circuit. The distance of separation between the algorithm and the attack was then increased progressively, and the resulting output fidelities are compared in FIG.~\ref{fig:separation}. Here we use fidelity rather than TVD as a measure of closeness between the target output and simulated noisy output, as the fidelity can be extracted directly from the density matrices used in simulation, rather than requiring costly sampling of the probability distribution. The results show a high level of variance in the effect of crosstalk on the device --- while some configurations with a separation of zero between the circuit and the attack show a significant deterioration in performance, some configurations are nearly unaffected. Increasing the radius of separation between circuits increases the average fidelity, and also reduces the variance between outcomes. This indicates that the radius of separation is unnecessarily heavy handed --- FIG.~\ref{fig:separation}(a)  shows how a significant fraction of the device must be left idle for even a small radius of separation, while FIG.~\ref{fig:separation}(b) shows that that for some configurations, only a small or zero separation is necessary to prevent interference. This implies that intelligent selection of qubits could allow circuits to be placed close to each other without crosstalk interference between them. To address these issues, a machine learning approach is developed below, which will scale to larger devices and automate the process of identifying good qubits.

\newcommand{\ibm}[2]{
    \def\radius{0.2}

    \foreach \x in {1,...,10} {
        \foreach \y in {0,2} {
            \draw[fill=qubit, draw=none] ({\x - \y / 2 + #1}, {\y + #2}) circle (\radius);
            \ifnum\x<9
                \draw[] ({\x + \radius + #1}, {\y + #2}) -- ({\x + 1 - \radius + #1}, {\y + #2});
            \fi
        }
    }
    \draw ({0 + #1 + \radius}, {2 + #2}) -- ({1 + #1 - \radius}, {2 + #2});
    \draw ({9 + #1 + \radius}, {0 + #2}) -- ({10 + #1 - \radius}, {0 + #2});

    \draw ({1 + #1}, {0 + #2 + \radius}) -- ({1 + #1}, {1 + #2 - \radius});
    \draw ({1 + #1}, {1 + #2 + \radius}) -- ({1 + #1}, {2 + #2 - \radius});
    \draw ({5 + #1}, {0 + #2 + \radius}) -- ({5 + #1}, {1 + #2 - \radius});
    \draw ({5 + #1}, {1 + #2 + \radius}) -- ({5 + #1}, {2 + #2 - \radius});
    \draw ({9 + #1}, {0 + #2 + \radius}) -- ({9 + #1}, {1 + #2 - \radius});
    \draw ({9 + #1}, {1 + #2 + \radius}) -- ({9 + #1}, {2 + #2 - \radius});

    \draw ({3 + #1}, {0 + #2 - \radius}) -- ({3 + #1}, {-1 + #2 + \radius});
    \draw ({3 + #1}, {2 + #2 + \radius}) -- ({3 + #1}, {3 + #2 - \radius});
    \draw ({7 + #1}, {0 + #2 - \radius}) -- ({7 + #1}, {-1 + #2 + \radius});
    \draw ({7 + #1}, {2 + #2 + \radius}) -- ({7 + #1}, {3 + #2 - \radius});

    \draw[fill=qubit, draw=none] ({1 + #1}, {(1 + #2)}) circle (\radius);
    \draw[fill=qubit, draw=none] ({9 + #1}, {(1 + #2)}) circle (\radius);
    \draw[fill=qubit, draw=none] ({3 + #1}, {(3 + #2)}) circle (\radius);
    \draw[fill=qubit, draw=none] ({3 + #1}, {(-1 + #2)}) circle (\radius);
    \draw[fill=qubit, draw=none] ({7 + #1}, {(3 + #2)}) circle (\radius);
    \draw[fill=qubit, draw=none] ({7 + #1}, {(-1 + #2)}) circle (\radius);

    \draw[fill=algo, draw=none] ({5 + #1}, {(0 + #2)}) circle (\radius);
    \draw[fill=algo, draw=none] ({5 + #1}, {(1 + #2)}) circle (\radius);
    \draw[fill=algo, draw=none] ({5 + #1}, {(2 + #2)}) circle (\radius);
}

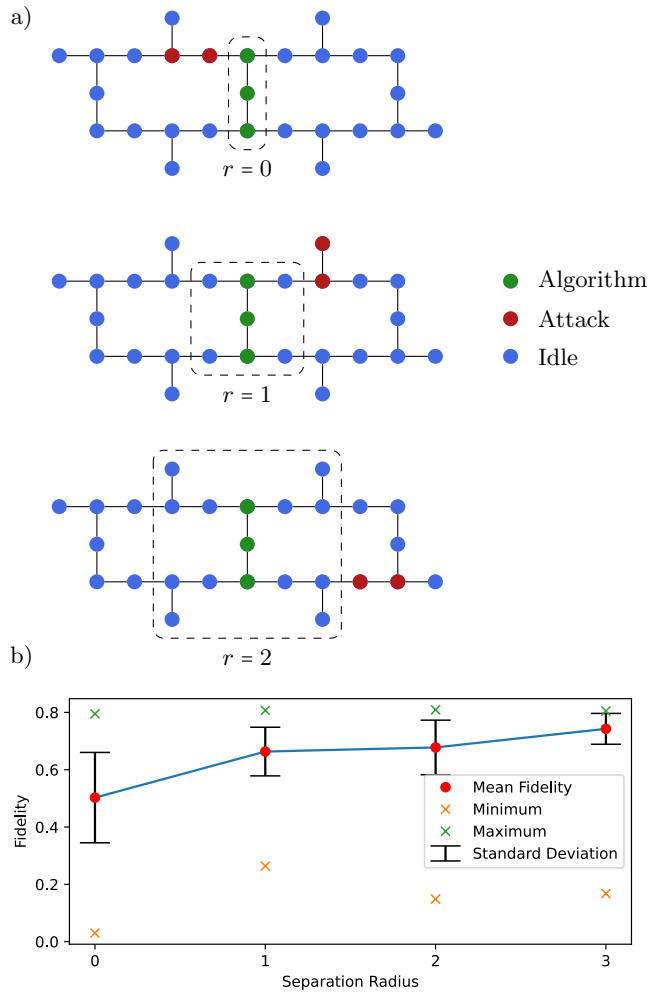
\begin{figure}
    \centering
    \begin{tikzpicture}[scale=0.5]
        \def\radius{0.2}
        \ibm{0}{0};
        \ibm{0}{-6};
        \ibm{0}{-12};
        
        \draw[fill=attack, draw=none] (3, 2) circle (\radius);
        \draw[fill=attack, draw=none] (4, 2) circle (\radius);
        
        \draw[fill=attack, draw=none] (7, -4) circle (\radius);
        \draw[fill=attack, draw=none] (7, -3) circle (\radius);
        
        \draw[fill=attack, draw=none] (8, -12) circle (\radius);
        \draw[fill=attack, draw=none] (9, -12) circle (\radius);
        
        \draw[rounded corners, dashed] (4.5, -0.5) rectangle (5.5, 2.5);
        \node at (5, -1) {$r=0$};
        \draw[rounded corners, dashed] (3.5, -3.5) rectangle (6.5, -6.5);
        \node at (5, -7) {$r=1$};
        \draw[rounded corners, dashed] (2.5, -8.5) rectangle (7.5, -13.5);
        \node at (5, -14) {$r=2$};

        \draw[fill=algo, draw=none] (12, -4) circle (\radius);
        \node[right] at (12.5, -4) {Algorithm};
        \draw[fill=attack, draw=none] (12, -5) circle (\radius);
        \node[right] at (12.5, -5) {Attack};
        \draw[fill=qubit, draw=none] (12, -6) circle (\radius);
        \node[right] at (12.5, -6) {Idle};
        
        \node at (7, -19.0) {\includegraphics[width=\linewidth]{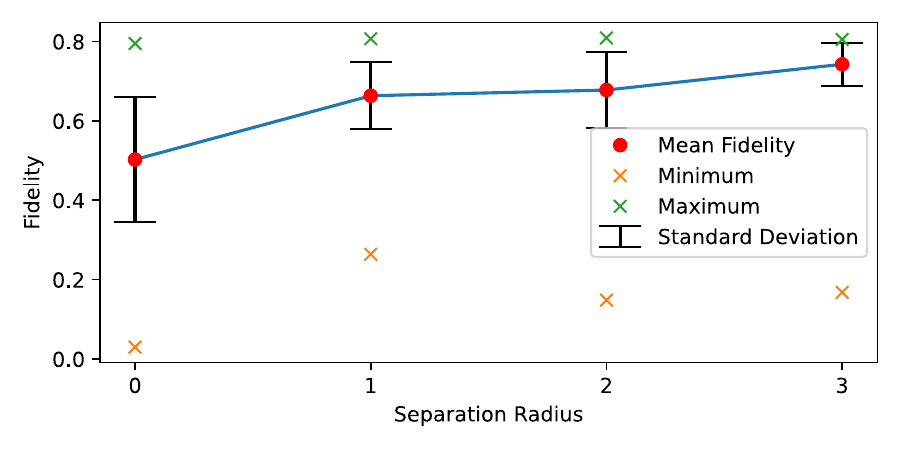}};
        
        \node at (-1, 3) {a)};
        \node at (-1, -14) {b)};
    \end{tikzpicture}
    \caption{ a) Circuits may be placed on the device with a separation radius between them. Here we place an circuit performing Grover's algorithm in green, while in red is a repeated CNOT attack. Attacks may be placed anywhere outside the radius of separation. b) When the circuits are directly adjacent, the output fidelities of Grover's algorithm are largely noise. If the separation radius between the circuits is increased, the output fidelities return towards the ideal distribution.}\label{fig:separation}
\end{figure}

\begin{figure*}[ht]
 \includegraphics[width=0.95\linewidth]{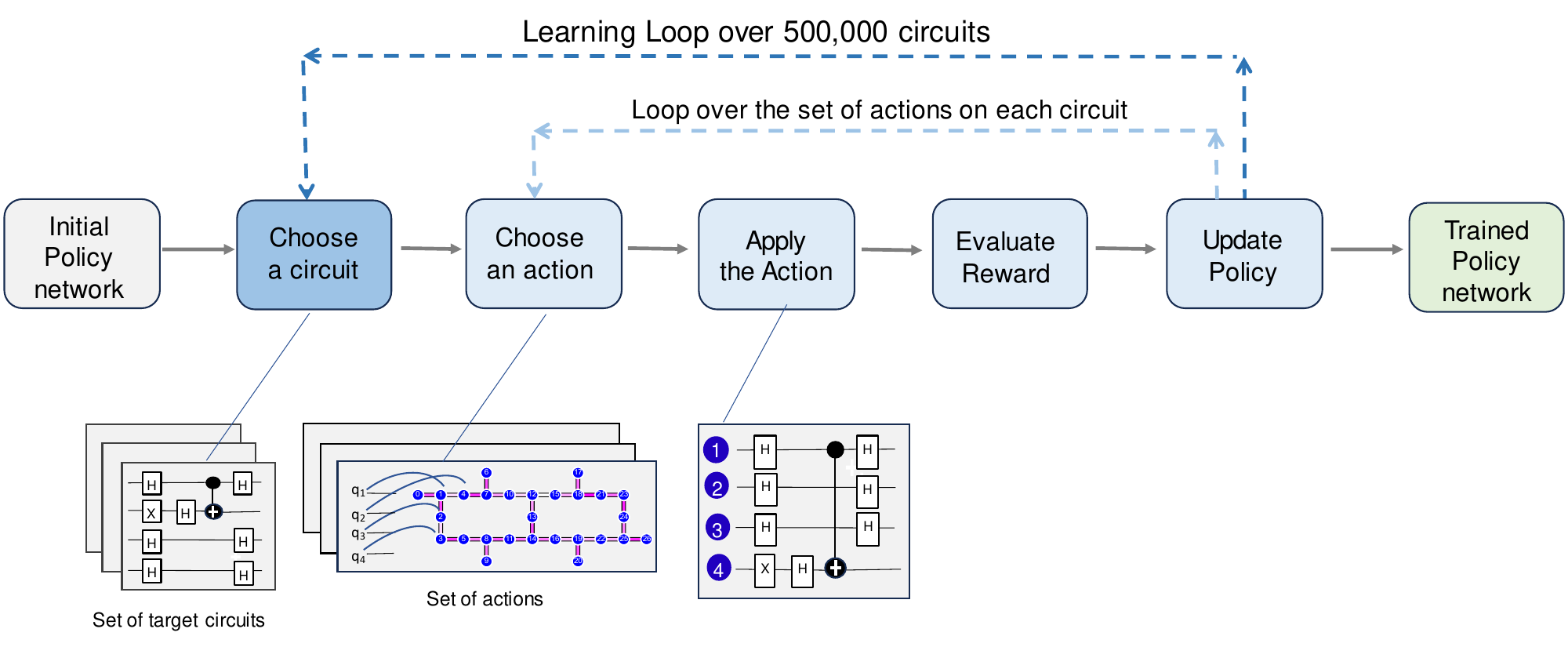}
    \caption{The flowchart shows the learning cycle for optimizing quantum circuit mapping. It begins with an initial, untrained policy network. In each iteration, the network picks a circuit, and for each circuit, it evaluates different mappings (actions). The number of actions is defined as a hyperparameter within the policy network. The efficacy of each action is assessed through a reward system, which informs the policy network. After training, the trained policy network can be applied to a new circuit, using its learned strategies to predict the optimal qubit mapping. Pseudocode describing the learning cycle is also shown in algorithm~\ref{alg:reinforcement}}
    \label{fig:Loop}
\end{figure*}

\begin{figure*}[ht]
\centering
    
    \raggedright % Aligns the label to the left
    \raisebox{\dimexpr\height-\baselineskip}[0pt][5pt]{\makebox[\textwidth][l]{\textbf{(a)}}}
    \centering
    \includegraphics[width=\textwidth]{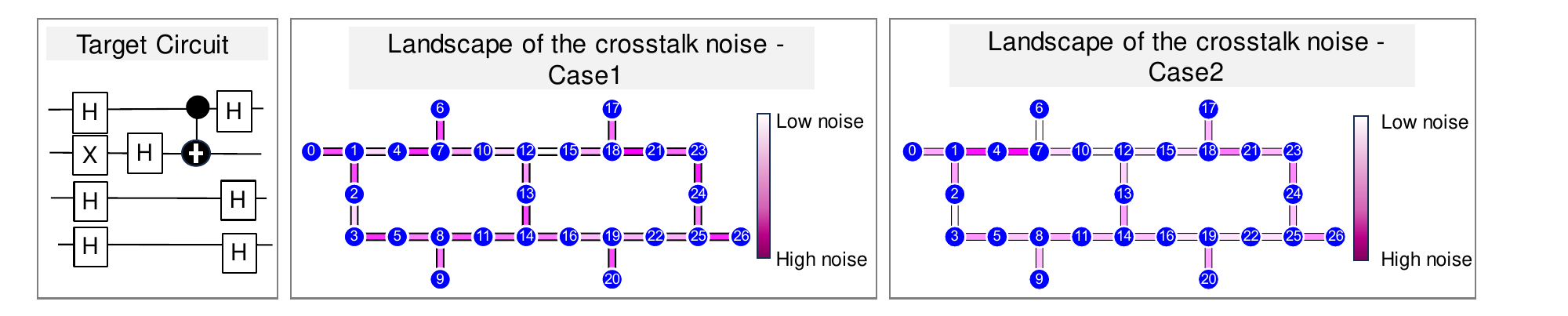}
    \label{fig:sub1}

\vspace{5pt}

    \raggedright
    \raisebox{\dimexpr\height-\baselineskip}[0pt][5pt]{\makebox[\textwidth][l]{\textbf{(b)}}}    \centering
    \includegraphics[width=\textwidth]{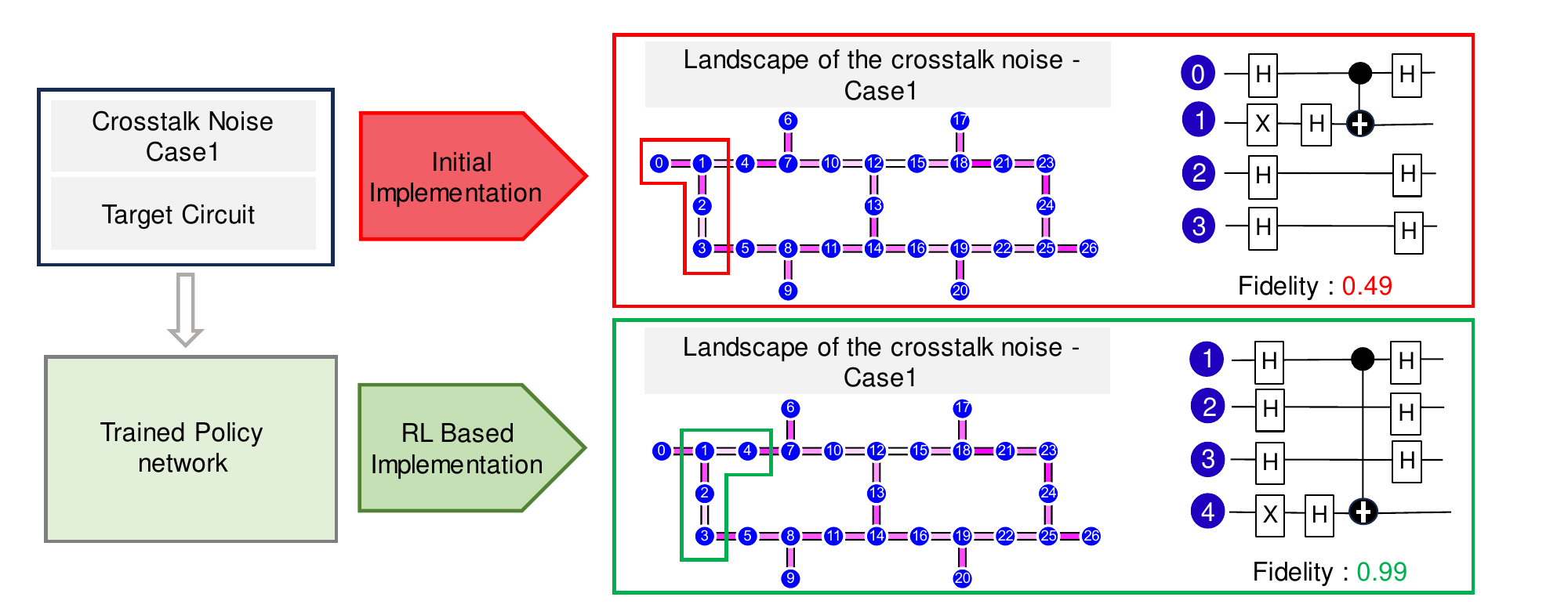}

    \label{fig:sub2}

\vspace{5pt}

    \raggedright
    \raisebox{\dimexpr\height-\baselineskip}[0pt][5pt]{\makebox[\textwidth][l]{\textbf{(c)}}}
    \centering
    \includegraphics[width=\textwidth]{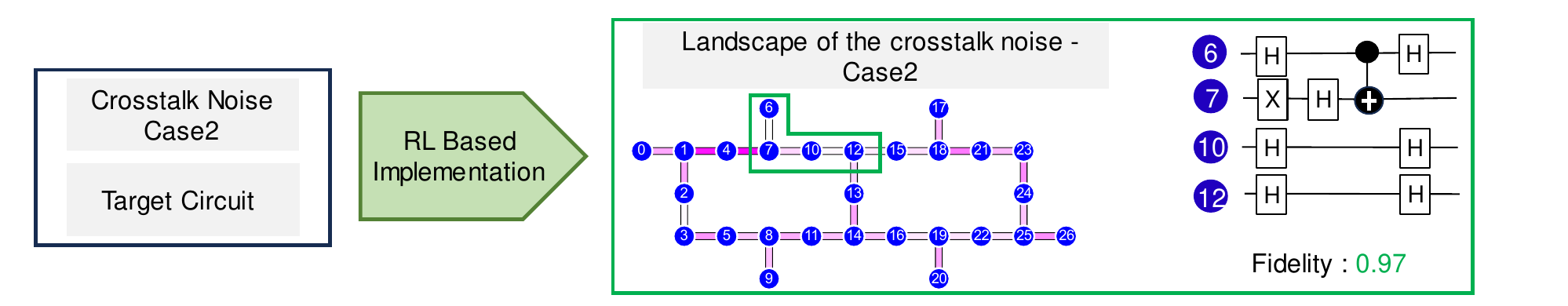}
    
\vspace{5pt}

    \raggedright
    \raisebox{\dimexpr\height-\baselineskip}[0pt][5pt]{\makebox[\textwidth][l]{\textbf{(d)}}}
    \centering
    \includegraphics[width=\textwidth]{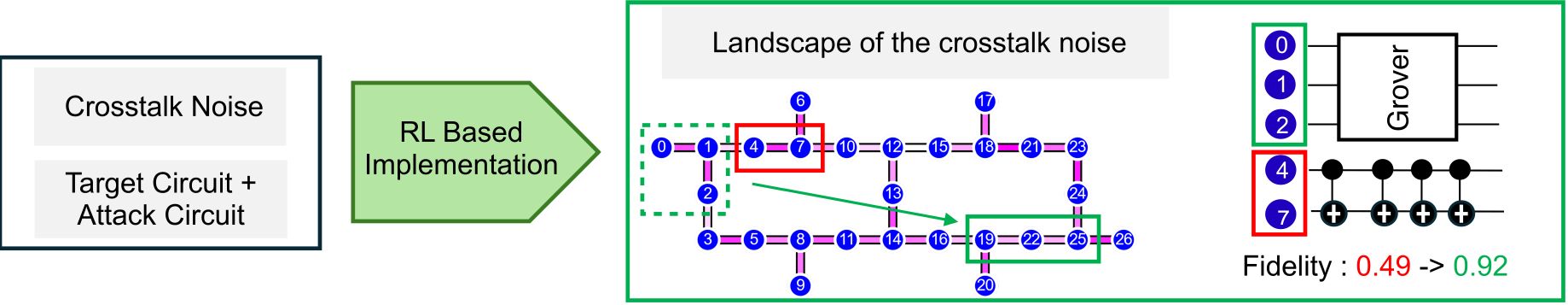}

\caption{Comparison of circuit mapping strategies in the presence of two different noise landscapes. (a) Displays the target circuit with two noise scenarios, Case1 depicts the crosstalk noise landscape benchmarked in section II, and Case2 a noise profile artificially altered for testing the reinforcement learning agent. The colouring of this figure is illustrative of the noise induced on adjacent qubits by performing that particular CNOT gate, calculated by taking the normalised average of the HSA components of the noise model.  Each CNOT may have a more or less detrimental effect on particular qubits. (b) A target circuit is introduced, and an initial mapping shows poor fidelity. In contrast, utilizing the reinforcement learning (RL) method, the circuit is strategically placed on the device, causing a significantly improved fidelity. A data set of 500,000 simulated circuits was used to train the RL agent. (c) Demonstrates RL's robustness in achieving optimal fidelity despite varied noise conditions in Case 2. (d) Shows how a victim circuit and a crosstalk attack may be positioned by the reinforcement learning algorithm to mitigate the effects of crosstalk noise. In this case the victim and attacker circuit are initially schedule adjacent to each other. The machine learning agent then relocates the victim circuit to prevent disruption by the attack circuit. The circuit used in this example is a 3-qubit instance of Grover's algorithm.}
\label{fig:Flowchart}
\end{figure*}

\subsection{Mapping Circuits with Reinforcement Learning}

As discussed in the introduction, crosstalk error arises from the unwanted effects of controls on nearby qubits. Consequently, the allocation of physical qubits for a given algorithm will influence the impact of crosstalk error. Thus, finding the optimal qubit allocation is a crucial task to achieve the optimal accuracy of computation. In this section, we utilize reinforcement learning (RL) to mitigate crosstalk noise by optimizing qubit allocation. The proposed method improves circuit fidelity by identifying and implementing practical mapping computations onto qubits that show reduced crosstalk errors.

While there are existing algorithms for scheduling one or multiple circuits on a noisy device to minimize errors~\cite{9407180},\cite{10.1117/12.3002854}, some of which include crosstalk~\cite{Niu_2023}, these algorithms only consider scheduling of arbitrary circuits. Our technique is used to schedule circuits which may be crosstalk attacks, with significantly more crosstalk effects than a neutral circuit that happens to be on the same device.

Our method uses RL, a process in which an algorithm iteratively refines its decision-making through feedback from a structured training regimen~\cite{quetschlich2023compiler}. In our specific context, the algorithm learns to map a target circuit  to physical qubits, taking into account the quantum computer's noise profile, particularly crosstalk, and identifying the mapping that minimizes errors. Previous studies~\cite{baum2021experimental, nautrup2019optimizing} have demonstrated the effectiveness of RL in reducing logical errors in the presence of Pauli noise channels. Our work extends RL techniques to reduce the rate of errors due to crosstalk noise. The proposed method will be discussed in more detail in the following section.

\subsubsection{Methodology and Definition of Framework Components}

Our framework aims to develop a trained policy network capable of effectively mapping quantum circuits to physical qubits, which is illustrated in FIG.~\ref{fig:Loop}. The main components of an RL method are \textit{action}, \textit{RL-state} and \textit{policy}. In our model these components are as follows.

The “action” in our RL model represents the allocation of qubits to physical qubits on the quantum computer, choosing how each qubit in a quantum circuit should be assigned to a physical qubit on the quantum computer. This means every action is essentially a permutation—a unique arrangement—where each qubit from the quantum circuit is mapped to one of the available physical qubits. These actions are constrained by the hardware’s topological limitations, dictating the possible interactions between qubits. The number of actions is a hyperparameter of our model. Central to our RL model’s framework is the “RL-state”, or observation, which represents the current configuration of the quantum circuit. The RL state includes factors such as the number of qubits and the sequence of quantum gates.

The policy network acts as the decision-making component within the reinforcement learning framework. Our policy is represented by a neural network with two layers, which through the use of gradient descent, learns to optimize the mapping of qubits to their physical counterparts. 
Through an iterative process shown in FIG.~\ref{fig:Loop}, or as pseudocode in algorithm~\ref{alg:reinforcement}, the neural network continuously learns by processing the current state of the quantum circuit and evaluating each mapping's effectiveness.

The effectiveness of each action is assessed through a designed reward function. The reward is calculated in terms of the fidelity of the executed circuit,
\[ \text{Reward} = \braket{\psi_\text{no noise}}{\psi_\text{noisy}}, \]
where circuits are executed on a simulator using the benchmarked device noise model described in section II.
This reward guides the policy network to identify the optimal mappings. Over time, the network learns to choose mappings that enhance circuit performance consistently.
After the policy network has completed a set of actions and received the corresponding feedback, it updates its policy using gradient descent. Once the policy is updated, the network proceeds to the next quantum circuit. The reinforcement learning algorithm used here is REINFORCE --- a Monte Carlo Policy Gradient method.
The dark blue loop shown in the FIG.~\ref{fig:Loop} encircles the set of target circuits, which are the training data for our model. In each loop a new quantum circuit is introduced to the policy network. The network processes this circuit to decide which action to take. Once an action is taken, the network reviews the results to learn from its decision. This review process is critical because it provides feedback that the policy network uses to refine its parameters. As a result, the policy network's ability to select the most rewarding actions is progressively enhanced through this iterative learning.

The training circuits used in this work consist of 5 qubit circuits on a randomly chosen subset of the simulated device. Circuits consisted of up to 20 gates chosen uniformly at random from the device's gate set.

Once training is complete, the trained policy network can receive any quantum circuit and efficiently predict its optimal mapping. This is a critical feature of the model, as it allows for practically applying the policy network's learned strategies to new, unseen quantum circuit configurations.

Another significance of our method is its adaptability to varying noise. Since noise landscape in quantum computers changes on daily basis adaptability to is essential. Therefore, the ability to update mappings without complete re-training is crucial. This process begins with a trained reinforcement learning model familiar with a specific noise environment. The model then undergoes partial re-training using data that reflects the latest noise profiles, which could influence qubit allocation. We fine-tune the model on a smaller dataset (20 \%  of the full dataset) to enable it to quickly adapt to the new noise conditions. The learning rate is systematically reduced every 10 episodes using a step decay plan. This gradual reduction helps refine the network gently, preventing drastic changes that could lead to overfitting of the new data.

\subsubsection{Results}
To achieve the objective of having a robust model, we need a large set of data. Note that covering every possible permutation of circuits is time consuming and impractical due to the massive variety of potential quantum circuit combinations. Despite this barrier, our data set is sufficient for our object as we discuss below.

The number model was trained on a dataset containing $500,000$ different circuits. These training circuits were randomly generated, with a uniform distribution of gates, and a uniform random number of up to 5 qubits and 20 gates chosen. We included various circuit configurations with different numbers of qubits, gates, and types of gates to ensure our dataset was comprehensive. It allows our model to learn how to handle a wide selection of quantum circuit setups effectively. A repository of these circuits is available upon request.

Examples of our trained model are illustrated in FIG.~\ref{fig:Flowchart}. In this figure, we compare the performance of the RL-based circuit mapping to that of an initial mapping. The initial mapping we refer to is a linear mapping approach where the first qubit is mapped onto the first physical qubit (qubit 0), the second qubit to the second physical qubit (qubit 1), and so on sequentially. This comparison highlights the effectiveness of RL-based circuit
mapping over initial mapping under various noise conditions. In a sample of 100 random test circuits, the reinforcement learning agent increased average output fidelity from 0.53 to 0.86.
FIG.~\ref{fig:Flowchart}(a)  presents the target circuit for implementation on a physical quantum computer. The study employs two distinct noise scenarios: Case$1$ is the intrinsic noise environment of an IBM quantum computer, while Case$2$ introduces an intentionally modified noise framework, where the noise between qubits 1 and 4 is intentionally increased. This modification was designed to test the robustness of our RL model.

Here, the fidelity is the probability of the circuit producing the correct output.
In contrast to the initial mapping, which causes low fidelity, the RL-based strategy in FIG.~\ref{fig:Flowchart}(b)  demonstrates a significant improvement in fidelity. The improvement in fidelity is achieved through the RL algorithm's strategic circuit placement, enhancing overall performance. FIG.~\ref{fig:Flowchart}(c)  further investigates the resilience of the RL method by introducing an artificial variation in the noise landscape, represented by Case$2$ in FIG.~\ref{fig:Flowchart}(a). Despite the altered conditions, the RL method consistently maintains high fidelity. This consistency underscores the method's adaptability to the noise.

To further test the performance of the RL mapping algorithm on a realistic example, we simulated a 3-qubit instance of Grover's algorithm in the presence of a crosstalk attack, similar to exeperiment in section~\ref{separation}. FIG~\ref{fig:Flowchart}(d) shows how the RL agent mapped Grover's algorithm in the presence of a crosstalk attack. We found that the RL agent was able to improve the fidelity of the algorithm from 0.49 to 0.92, demonstrating the versatility of a RL approach. Finally, we compare the RL agent to a naive separation radius. Table~\ref{separation_table} shows that the RL agent is able to select the optimal separation configuration on the device, while in table~\ref{grover_table} we force the RL agent to position the Grover's algorithm circuit and crosstalk attack adjacent to each other. The results show that given a fixed separation radius, the RL agent is able to select the optimal mapping of qubits, whereas a simple separation radius may pick an inferior configuration, as it is unaware of the noise landscape.

\begin{table}[h]
    \begin{tabular}{c c c}
        \toprule
        Algorithm Qubits &  Separation Radius & Fidelity \\
        \midrule
        0, 1, 2 & 0 & 0.49 \\
        3, 5, 8 & 2 & 0.59 \\
        11, 14, 16 & 3 & 0.68 \\
        21, 23, 24 & 4 & 0.81 \\
        \textbf{19, 22, 25} & 5 & \textbf{0.92} \\
        & 
    \end{tabular}
    \caption{The performance of Grover's algorithm when positioned in various configurations on the device. In all cases, the crosstalk attack is a repeated CNOT on qubits 4 and 7. Increasing the radius of separation between the circuit and the crosstalk attack improves the performance, however the RL agent is able to immediately identify the optimal mapping, highlighted in bold.}\label{separation_table}
\end{table}

\begin{table}[h]
    \begin{tabular}{c c}
        \toprule
        Algorithm Qubits & Fidelity \\
        \midrule
        \textbf{7, 10, 12} & \textbf{0.65} \\
        5, 8, 11 & 0.57 \\
        16, 19, 22 & 0.49 \\
        & 
    \end{tabular}
    \caption{The performance of Grover's algorithm when positioned in various configurations on the device. In all cases, the crosstalk attack is a repeated CNOT on qubits 13 and 14. All allowed configurations have the circuit adjacent to the crosstalk attack, i.e. separation radius = 0. While the initially chosen placement performs poorly, the RL agent is able to identify the configuration with the highest output fidelity, shown in bold. The separation radius defence is unable to identify the superior configuration at a fixed separation radius.}\label{grover_table}
\end{table}

Scaling this technique to larger circuits will require including circuits over larger numbers of qubits in the training dataset. This quickly becomes intractable to simulate classically, so shots must be run on a physical quantum device, using circuits where the expected output is already known. Computing training circuits on a real device and updating the learning agent are tasks which must be fast enough to repeat regularly, without significantly reducing the amount of time the device has available to perform computations.

Generalising these results to other hardware architectures is relatively straightforward. As 2-qubit gates are the main source of crosstalk, any 2-qubit operation which is related to the CNOT gate by single qubit operations may use our protocol unchanged. This is true for many quantum computing platforms.

\begin{algorithm}[H]
    \caption{Reinforcement training loop}\label{alg:reinforcement}
    \begin{algorithmic}
        \State Policy Network $\gets$ initial
        \For{circuit $\in$ training data}
            \For{mapping $\in$ actions}
                \State mapped circuit $\gets$ mapping(circuit)
                \State fidelity $\gets$ reward(mapped circuit)
                \State Policy Network $\gets$ update(Policy network, fidelity)
            \EndFor
        \EndFor
    \end{algorithmic}
\end{algorithm}

\begin{figure*}
    \centering
    \includegraphics[width=\linewidth]{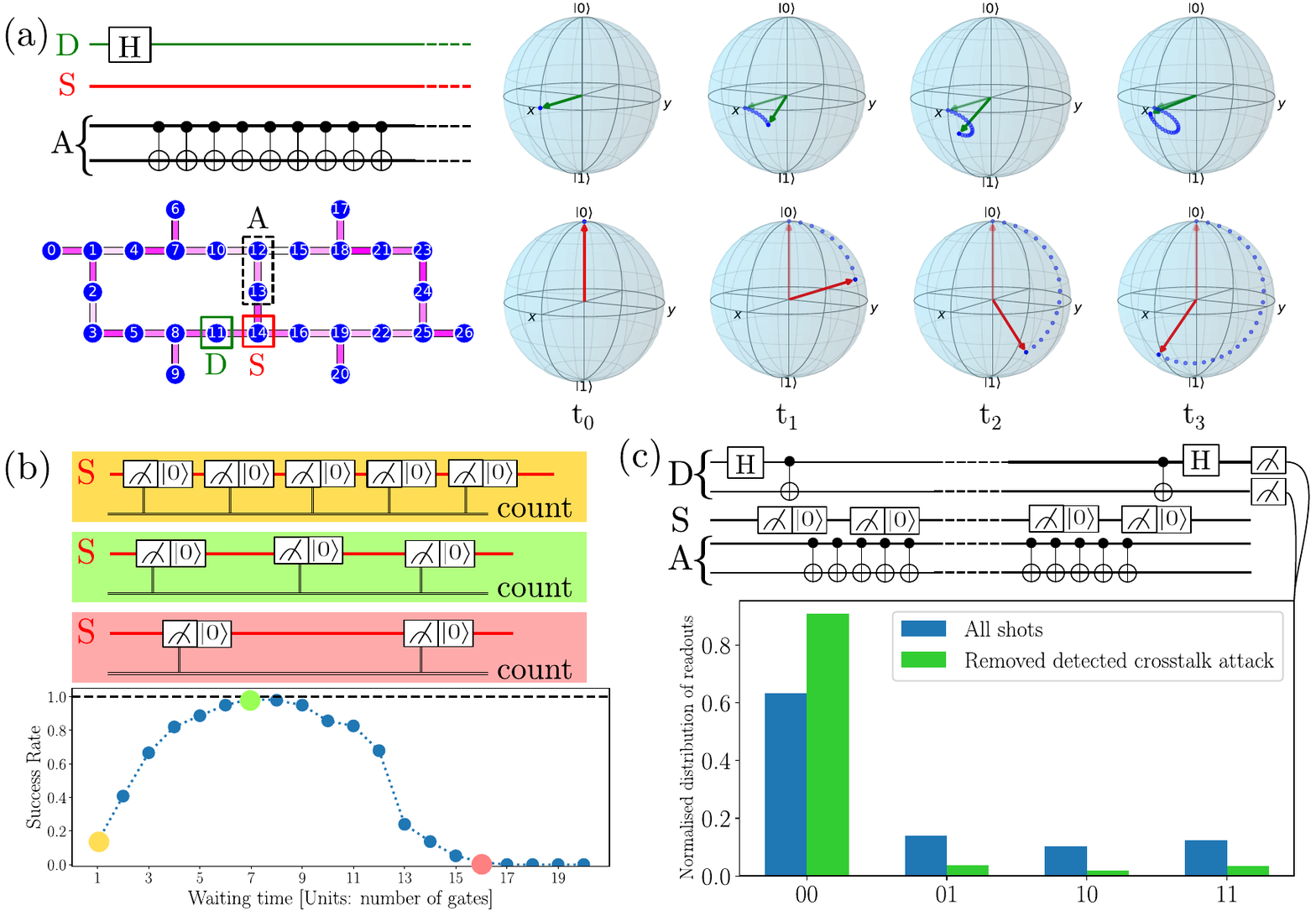}
    \caption{(a) Demonstration of correlated dynamics of data qubit and spectator qubit under the effect of crosstalk attack. Here, qubit $11$ of the IBM device, is chosen as data qubit and initialised at $|+\rangle$ using a Hadamard gate. Qubit $14$ plays the role of the spectator and its initial state is $|0\rangle$. Crosstalk attack is performed via a series of CNOT gates on qubits $12$ and $13$. Bloch spheres of data qubit (top row) and spectator qubit (bottom row) at different times demonstrate the correlated dynamics between the spectator and the data qubit. (b) Schematic for finding the optimum waiting time to detect the presence of the crosstalk attack. If the waiting time is chosen to be too short or too long, the spectator qubit would not detect the presence of the crosstalk attack. (c) By removing shots with detected crosstalk we mitigate the effect of the crosstalk. }
    \label{fig:Spectator}
\end{figure*}

\subsection{Crosstalk Attack detection via a Spectator Qubit}

The use of Spectator Qubit (SQ) in quantum control is a novel idea that has been studied theoretically \cite{Tonekaboni2023, Song2023} and experimentally \cite{singh2023mid}. A SQ refers to a qubit that coexists inside the same physical space as a data qubit (DQ). The SQ should be positioned in close proximity to the data qubit in order to effectively capture the same noise. However, it should also be positioned at a sufficient distance to avoid any potential interaction with the data qubit. The objective is to perform frequent measurements on the SQ in order to gain knowledge about the effect of noise on the data qubits. Subsequently, equipped with the acquired knowledge, we have the choice of correcting the imprecision in the data qubit to a certain degree or declaring that the data qubit has been considerably affected, resulting in the loss of its quantum information. Despite recent research on the SQ idea, its direct use on quantum computing devices has not been demonstrated. Historically, the main obstacle to implementing SQ on quantum computing devices is the lack of multiple measurements, resets, and feedback loops inside a running circuit in current devices. As these features were absent on the device we benchmarked, we instead used the results of the IBM quantum device characterization from Sec.~\ref{sec:characterization} to replicate the SQ notion for detecting the presence of nearby crosstalk attacks.

\subsubsection{Dynamical correlation between the spectator and the data qubits}
The key assumption underlying the SQ concept is that SQ and DQ exhibit correlated dynamics when subjected to noise. In previous studies on SQ, this correlation has been artificially introduced either by choosing a similar Hamiltonian for both SQ and DQ in theoretical studies \cite{Tonekaboni2023, Song2023} or by using engineered noise in an experimental setting \cite{singh2023mid}.

In contrast, we intend to implement the SQ concept on an IBM device without engineering any noise or Hamiltonian. Thus, in the first stage, we analyse the correlated dynamics of two qubits, one as the DQ and one as the SQ, under the influence of a crosstalk attack. To accomplish this, we simulate the dynamics of the SQ and DQ for various settings on the 27-qubit IBMQ using the crosstalk noise model characterised from the real $27$-qubit device described in Sec.~\ref{sec:characterization}.

FIG.~\ref{fig:Spectator}(a) shows an example of the aforementioned simulation. Here, we choose qubits $11$ and $14$ as the DQ and the SQ respectively, as depicted by the solid green(red) square for DQ(SQ) on the topology of the device. Also, we choose qubits $12$ and $13$ as attack qubits (AQ) depicted via dash black squared on the topology. Furthermore, the quantum circuit in FIG.~\ref{fig:Spectator}(a) shows our approach to simulating the dynamics of the SQ and DQ under the effect of a crosstalk attack. At time $t=t_0$, we initialize the DQ in $\ket{+} = (\ket{0} + \ket{1})/\sqrt{2}$ via a Hadamard gate and leave the SQ in $\ket{0}$ state. We then apply a series of CNOT gates on AQ. 

Our simulation confirms the correlated dynamics of the SQ and DQ. In FIG.~\ref{fig:Spectator}(a), we demonstrate this correlated dynamics by picturing the state of the DQ and SQ at different times via Bloch sphere. The top row shows the state of DQ at different times, while the bottom row shows the state of SQ. The states of the SQ and the DQ at $t=t_0$ are $\ket{0}$ and $\ket{+}$ respectively, as expected. As time passes, at $t_1, t_2, $ and $t_3$, we observe that the states of both SQ and DQ rotate in the same direction (the rotation trajectories are shown by points). The synchronised rotations are caused by the crosstalk attack which confirms that SQ and DQ's dynamics are correlated.

We emphasise here that this correlated dynamics is simulated using the data from the characterization of the real IBM device (see Section II), which represents the crosstalk effect in the real device and is not engineered correlated noise. 

\subsubsection{Measurement Strategy for the Spectator Qubit}
We showed the correlated dynamics of SQ and DQ. The idea is to gather information about noise by measuring the SQ. However, measuring the spectator qubit causes its state to be projected to either $\ket{0}$ or $\ket{1}$, so it might not be a useful SQ after a measurement. To overcome this, following a similar setup as in \cite{Tonekaboni2023}, we repeatedly measure the SQ, record the result, and  reset the SQ to $\ket{0}$ state; we then repeat the measurement/reset process many times. As in ~\cite{Tonekaboni2023}, the waiting time $\tau$ between the measurement/reset cycle is,
\begin{equation}
    \tau = \frac{1}{K}
\end{equation}
where $K$ is the sensitivity of the spectator qubit to noise. We note that although frequent measurement on the SQ might cause additional noise in the device, we expect this effect to be negligible compared to CNOT gates. This is because measurements happen less often than attack CNOT gates and only involve a single qubit, whereas two-qubit gates have more significant crosstalk noise. The challenge is to find out how frequently we must measure and reset the SQ. If we measure the SQ often, it will not have enough time to be impacted by the crosstalk attack; if we wait too long before measuring the SQ, its state may rotate back to its initial value or be influenced by other noise rather than the crosstalk that we are interested in. 

To address the above question, we consider a scenario where we would like to detect the presence of a crosstalk attack. We use a similar setup as before, i.e., qubit $14$ as SQ and qubits $12$ and $13$ as AQ (note that we do not need DQ in this scenario). The goal is to identify if a crosstalk attack is present by measuring the SQ. To do that, we measure the SQ repeatedly and reset it after each measurement. In the absence of crosstalk, the spectator qubit stays at state $\ket{0}$ so the measurement result would always be `0'. Thus, if the measurement results in `1', it would be a \textit{flag} for the presence of an attack. To continue, we count the number of flags and store them in $f$; then, if the number of flags is greater than a predefined threshold $f_0$, we indicate that the shot was under attack. The threshold $f_0$ is included to distinguish existing of any other noise that might present. Note that, in an ideal simulation where the noise is due to only crosstalk attacks, $f_0=0$, but if we consider other possible noise, especially measurement noise, we may consider $f_0>0$ to be more accurate in deciding the presence of an attack.

Furthermore, we run $N$ shots of the setup (all shots are under attack) and we define a success rate $\eta=N_d/N$ where $N_d$ is number of detected shots under attack.
We repeat the procedure for different waiting time and calculate the success rate $\eta$ for each waiting time to find the optimum waiting time. In our simulation we use $N=1000$ shots and waiting time defined in units of number of CNOT gates. We also include a $1\%$ chance of measurement error in SQ readout; considering this and the number of shots, we chose $f_0 = 11$, which is slightly greater than $1\%$ of the total number of shots. In FIG.~\ref{fig:Spectator}(b), we illustrated three highlighted scenarios: measuring the SQ too frequently (orange), optimal waiting time (green) and long waiting time (red). We note that in our simulation, we assumed that gate and measurement times are the same. Since, in IBM devices, the measurement time is approximately twice as long as the CNOT gate time, the simulation remains reliable as long as the number of CNOT gates is more than a few (two or three) which is the case in our setup and results. The plot in FIG.~\ref{fig:Spectator}(b) shows our simulation result. As we expected, if we measure the spectator qubit after only one gate, its state has not evolved enough to detect the attack. On the other hand, if we wait too long, we see the state rotates back to $\ket{0}$ and we get low success rate. Our simulation shows that optimum waiting time (the time between two measure/reset pairs on SQ) for our setup is approximately $7$ CNOT gates. We note that the waiting time that we found is based on the aforementioned setup and it depends on the rate of the CNOT gates, the location of the attack qubits, and the crosstalk noise values. \\

\subsubsection{Detection of the crosstalk attack}

Now that we have a protocol to detect the presence of a crosstalk attack, we can use it to post-select our shots. To demonstrate this idea, we simulate the following scenario: Consider two data qubits initialized in an entangled state, which after some time we would like to use. Also, consider that some but not all of the shots are under crosstalk attack. Although we do not know which shots are under attack, we use the spectator qubit to detect the impacted shots and discard them, i.e., post select on the shots that are not under attack. In our simulation, at the end of the circuit, we disentangle the qubits before measurement, so the measurement result of the data qubits should be $00$. {Also, in the simulation, we choose that $20\%$ of the shots are under crosstalk attack. This percentage is arbitrary, but because we are discarding the attacked shots, we choose a relatively small percentage (below $50\%$). FIG.~\ref{fig:Spectator}(c) shows our circuit as well as the normalized histogram of the final count of the data qubits. Blue represents all shots, while green shows the post-selected shots. From the histogram, it is obvious we have successfully discarded many of the impacted shots.

We pause here to note that the introduction of spectator qubits introduces a separation between data qubits on a device, and we earlier showed that this improves resilience to crosstalk noise. In this section, the result of simply separating the data and attack qubits are shown in FIG~\ref{fig:Spectator}(c) as ``All shots'', as this is the case where the spectator qubit is left to idle.

Although we successfully flagged the presence of the attack, discarding the shots is wasteful. Correcting the impacted shots using the SQ has been done in \cite{Tonekaboni2023, Song2023} for dephasing noise. Correcting the qubit state that is impacted by a crosstalk attack is more complicated, and we leave it for our future work. Another avenue for future work is to consider the crosstalk effect of SQ measurement/reset on DQ. Although, as mentioned before, we do not expect it to be a concerning issue, we plan to incorporate this into our model in future work. This will enable us to achieve a more realistic simulation that can be implemented on real devices. Finally, for a larger circuit, we probably need more than one spectator qubit. So, optimizing the multi-spectator qubit is another objective for future work. 

\section{Conclusion and Discussion}\label{sec:conclusion}
Crosstalk noise is a serious issue for near term and future large scale quantum devices. In this paper we have demonstrated the adverse effects of crosstalk on a circuit, particularly in a device shared by many users. We developed a full map of the crosstalk on a real IBM device, and used this to create a high fidelity simulator for the device.

We then developed a number of techniques to detect and/or mitigate the effects of a hostile crosstalk attack on a shared device. The first technique, circuits separation, was effective, but sees a substantial amount of the device left to idle, while scaling poorly to devices with large numbers of qubits. We then developed a reinforcement learning model to learn the noise on the device, and placed circuits in such a way that minimises the potential for crosstalk interference on the device. The final technique was based on the idea of spectator qubits, which have already been used for detecting and mitigating general noise models. We applied the spectator qubit technique to the problem of crosstalk noise, and found that it was highly effective at detecting the presence of a crosstalk attack. Further work on the spectator qubit technique includes correcting for a crosstalk induced error, rather than simply discarding erronous shots. It is also necessary to further explore the optimal placement of spectator qubits around a circuit, and the performance of spectator qubits with different attack circuits and varying attack strength.

The crosstalk attacks studied here, and elsewhere in the literature, are simple repeated CNOTs, designed to maximise the disrupting effect of crosstalk. It may be possible to create an attack which disrupts a computation in a targeted way, for example, by causing Grover's algorithm to select an incorrect answer with high confidence. Future work can explore such possibility and develop methods to detect and mitigate such scenarios.

There is the potential to flip the crosstalk attacks --- and instead use crosstalk to extract information. Quantum algorithms require classical data, which may be sensitive. This data is encoded in state preparation circuits, and a third party may be able to derive some knowledge of the secret data through exploiting crosstalk. While challenging, the potential impact of this could be significant, justifying further work.

Finally, the crosstalk benchmarking described here can be applied to a variety of hardware architectures, such as silicon qubits, trapped ions, or others. These are very different physical systems, which will have very different properties. It is important to properly explore the effect of crosstalk on other hardware, to see how bad (or not) it is. This work does not directly implement our crosstalk mitigation strategies on a real device, as the device benchmarked and used in simulation is no longer available for use. However based on the qualitative description of real device noise captured in our model, we believe that our results provides useful insights which are applicable to realistic hardware environments. In summary, we have laid the foundation to exploit crosstalk noise as a mechanism to attack and disrupt quantum computations in a shared environment. We have also reported mitigation strategies and demonstrated their working on proof of concept examples. In near to medium term era when quantum computing is anticipated to be based on shared environment through cloud platforms, securing it will be of paramount importance in particular for security sensitive data and computations.

\section*{Acknowledgment}
BH acknowledges the support of
the CSIRO Research Training Program
Scholarship. The work was partially
supported by the Australian Army through Quantum
Technology Challenge 2023. The computational resources
were provided by the National Computing Infrastructure
(NCI) and Pawsey Supercomputing Center through National
Computational Merit Allocation Scheme (NCMAS). The research was supported by the University of Melbourne
through the establishment of the IBM Quantum
Network Hub at the University. \\
\textbf{Data availability:} The data that support the find-
ings of this study are available within the article. \\
\textbf{Competing financial interests:} The authors declare no
competing financial or non-financial interests.

\bibliography{main}

\onecolumngrid
\appendix

\section{Total Variational Distance}\label{app:tvd}
For two probability distributions, $P$, $Q$ over the same space of outcomes $\Omega$, the Total Variational Distance (TVD) is the maximum difference between the distributions over all events,

\begin{equation} 
    \text{TVD}(P, Q) = \text{sup}_{E \in \Omega} |P(E) - Q(E)|
\end{equation}

Classical TVD is equivalent to the trace distance in a quantum system, and from the Fuchs–van de Graaf inequalities and Uhlmann's theorem, it can be shown that the trace distance $D$ (and hence TVD) can be related to state fidelity $F$ by~\cite{fuchs1998cryptographicdistinguishabilitymeasuresquantum}~\cite{UHLMANN1976273},

\begin{equation}
    1 - \sqrt{F(\rho, \sigma)} \leq D(\rho, \sigma) \leq \sqrt{1 - F(\rho, \sigma)}
\end{equation}

The TVD measures a distance between two distributions --- so when comparing the output fidelities of various simulators, a reference distribution is required. Previous work benchmarking the performance of a crosstalk simulator~\cite{ash-_saki_experimental_2020} has chosen to use an ideal noise-free simulation as the reference distribution, with the crosstalk simulator and real device measured relative to this. However, this may be misleading, as the real device and the crosstalk simulator may be a similar distance away from an ideal simulation, but not close to each other at all. As such, in this paper we use the real device probability distribution as reference, and so a better simulation of that real device has a lower TVD. For a circuit with real counts $c_d$, and simulated counts $c_s$, we find that the distance is,

\begin{equation}
    \text{dist}(c_s) = \text{TVD}(c_s, c_d)
\end{equation}

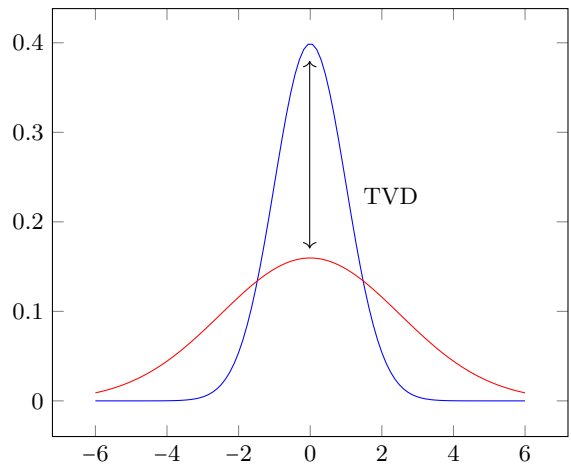
\begin{figure}
    \begin{tikzpicture}
        \begin{axis}
            \addplot[domain=-6:6,samples=100,color=blue] {1 / (2 * pi)^0.5 * e^(-x^2 / 2)};
            \addplot[domain=-6:6,samples=100,color=red] {0.4 / (2 * pi )^0.5 * e^(-(0.4*x)^2 / 2)};
        \end{axis}
        \draw[<->] (3.42, 2.5) -- (3.42, 5);
        \node at (4.5, 3.2) {TVD};
    \end{tikzpicture}
    \caption{\raggedright The total variational distance is the greatest distance between two probability distributions.}
\end{figure}

\section{Benchmarking Circuits}\label{app:circ}
Below are circuit diagrams for the circuits used in benchmarking the crosstalk simulator used in section~\ref{sec:benchmarking}.

\subsection{syn\_n}
This is a synthetic circuit where two idle qubits are prepared into a superposition. 
\begin{equation*}
    \begin{quantikz}
        & \lstick{\ket{0^{\otimes 2}}} & \gate{H} & \qw{} & \qw{} & \qw{} & \qw{} & \qw{} &
        \gate{H} &\meter{} \\
        & \lstick{$q_a$} & \qw{} & \ctrl{1} & \qw{} & \ldots\ & & \ctrl{1} & \qw{} & \qw{} \\
        & \lstick{$q_b$} & \qw{} & \targ{} & \qw{} & \ldots\ & & \targ{} & \qw{} & \qw{} \\
        & & & & & \underbrace{}_{n\text{ times}}
    \end{quantikz}
\end{equation*}
The CNOT gate is repeated $n$ times.

\subsection{toffoli}
This is a simple repeated Toffoli gate test. The Toffoli gates are decomposed into 1-qubit gates and the 2-qubit CNOT gate.
\begin{equation*}
    \begin{quantikz}
        & \lstick{$q_a$} & \gate{X} & \gate{H} & \ctrl{2} & \qw{} & \ldots\ & & \ctrl{2} & \gate{H} & \meter{} \\
        & \lstick{$q_b$} & \gate{X} & \gate{H} & \ctrl{1} & \qw{} & \ldots\ & & \ctrl{1} & \gate{H} & \meter{} \\
        & \lstick{$q_c$} & \qw{} & \gate{H} & \targ{} & \qw{} & \ldots\ & & \targ{} & \gate{H} & \meter{} \\
        & & & & & & \underbrace{}_{n\text{ times}}
    \end{quantikz}
\end{equation*}
The Toffoli gate is repeated $n$ times.

\subsection{grover}
The grover circuit is a simple 3-qubit Grover's algorithm circuit with two iterations.
\begin{equation*}
    \begin{quantikz}
        \ket{0} & \gate[3]{\text{Oracle}} & \gate[3]{\text{Diffusion Operator}} & \gate[3]{\text{Oracle}} & \gate[3]{\text{Diffusion Operator}} & \meter{} \\
        \ket{0} & \ghost{\text{Oracle}} & \ghost{\text{Diffusion Operator}} & \ghost{\text{Oracle}} & \ghost{\text{Diffusion Operator}} & \meter{} \\
        \ket{0} & \ghost{\text{Oracle}} & \ghost{\text{Diffusion Operator}} & \ghost{\text{Oracle}} & \ghost{\text{Diffusion Operator}} & \meter{} \\
    \end{quantikz}
\end{equation*}

\subsection{attack}
The attack test circuit is identical to the Grover's algorithm circuit, with the addition of a crosstalk attack on adjacent qubits;
\begin{equation*}
    \begin{quantikz}
        \lstick{\ket{0}} & \gate[3]{\text{Oracle}} & \gate[3]{\text{Diffusion Operator}} & \gate[3]{\text{Oracle}} & \gate[3]{\text{Diffusion Operator}} & \meter{} \\
        \lstick{\ket{0}} & \ghost{\text{Oracle}} & \ghost{\text{Diffusion Operator}} & \ghost{\text{Oracle}} & \ghost{\text{Diffusion Operator}} & \meter{} \\
        \lstick{\ket{0}} & \ghost{\text{Oracle}} & \ghost{\text{Diffusion Operator}} & \ghost{\text{Oracle}} & \ghost{\text{Diffusion Operator}} & \meter{} \\
        \lstick{} & \ctrl{1} & \ctrl{1} & \ctrl{1} & \ctrl{1} & \qw{} \\
        \lstick{} & \targ{} & \targ{} & \targ{} & \targ{} & \qw{} \\
    \end{quantikz}
\end{equation*}
CNOT gates are repeated during the execution of the circuit.

\section{Device Parameters}\label{app:device-params}
\begin{table}[!ht]
    \makebox[\textwidth][c]{
    \resizebox{1.1\textwidth}{!}{
    \begin{tabular}{|l|l|l|l|l|l|l|l|l|l|l|l|l|l|l|l|l|l|l|l|l|l|l|l|l|l|l|l|}
    \hline
        ~ & $q_{0}$ & $q_{1}$ & $q_{2}$ & $q_{3}$ & $q_{4}$ & $q_{5}$ & $q_{6}$ & $q_{7}$ & $q_{8}$ & $q_{9}$ & $q_{10}$ & $q_{11}$ & $q_{12}$ & $q_{13}$ & $q_{14}$ & $q_{15}$ & $q_{16}$ & $q_{17}$ & $q_{18}$ & $q_{19}$ & $q_{20}$ & $q_{21}$ & $q_{22}$ & $q_{23}$ & $q_{24}$ & $q_{25}$ & $q_{26}$ \\ \hline
        Frequency (GHz) & 5.04 & 5.16 & 5.26 & 5.10 & 5.07 & 5.21 & 5.02 & 4.92 & 5.03 & 4.87 & 4.82 & 5.16 & 4.72 & 4.96 & 5.05 & 4.92 & 4.88 & 5.22 & 4.97 & 5.00 & 5.10 & 4.84 & 4.92 & 4.92 & 4.99 & 4.81 & 5.02 \\ \hline
        $\text{T}_1 (\mu s)$ & 218.8 & 194.7 & 222.7 & 186.1 & 189.8 & 154.8 & 73.6 & 291.0 & 235.8 & 181.0 & 167.5 & 146.1 & 289.3 & 79.7 & 261.5 & 204.2 & 187.2 & 58.1 & 183.7 & 268.5 & 67.2 & 200.2 & 197.4 & 144.4 & 206.6 & 176.0 & 115.1 \\ \hline
        $\text{T}_2 (\mu s)$ & 190.7 & 124.2 & 292.7 & 28.4 & 14.4 & 203.4 & 201.0 & 162.8 & 315.4 & 110.2 & 124.4 & 144.1 & 305.9 & 21.5 & 26.7 & 41.8 & 230.8 & 44.0 & 188.5 & 194.4 & 44.2 & 34.2 & 98.7 & 245.5 & 34.4 & 101.6 & 22.4 \\ \hline
        $\text{CX}_{q_{0}; q_n}$ error (\%) & ~ & 0.8 & ~ & ~ & ~ & ~ & ~ & ~ & ~ & ~ & ~ & ~ & ~ & ~ & ~ & ~ & ~ & ~ & ~ & ~ & ~ & ~ & ~ & ~ & ~ & ~ & ~ \\ \hline
        $\text{CX}_{q_{1}; q_n}$ error (\%) & 0.8 & ~ & 0.3 & ~ & 0.6 & ~ & ~ & ~ & ~ & ~ & ~ & ~ & ~ & ~ & ~ & ~ & ~ & ~ & ~ & ~ & ~ & ~ & ~ & ~ & ~ & ~ & ~ \\ \hline
        $\text{CX}_{q_{2}; q_n}$ error (\%) & ~ & 0.3 & ~ & 0.7 & ~ & ~ & ~ & ~ & ~ & ~ & ~ & ~ & ~ & ~ & ~ & ~ & ~ & ~ & ~ & ~ & ~ & ~ & ~ & ~ & ~ & ~ & ~ \\ \hline
        $\text{CX}_{q_{3}; q_n}$ error (\%) & ~ & ~ & 0.7 & ~ & ~ & 0.6 & ~ & ~ & ~ & ~ & ~ & ~ & ~ & ~ & ~ & ~ & ~ & ~ & ~ & ~ & ~ & ~ & ~ & ~ & ~ & ~ & ~ \\ \hline
        $\text{CX}_{q_{4}; q_n}$ error (\%) & ~ & 0.6 & ~ & ~ & ~ & ~ & ~ & 1.3 & ~ & ~ & ~ & ~ & ~ & ~ & ~ & ~ & ~ & ~ & ~ & ~ & ~ & ~ & ~ & ~ & ~ & ~ & ~ \\ \hline
        $\text{CX}_{q_{5}; q_n}$ error (\%) & ~ & ~ & ~ & 0.6 & ~ & ~ & ~ & ~ & 100.0 & ~ & ~ & ~ & ~ & ~ & ~ & ~ & ~ & ~ & ~ & ~ & ~ & ~ & ~ & ~ & ~ & ~ & ~ \\ \hline
        $\text{CX}_{q_{6}; q_n}$ error (\%) & ~ & ~ & ~ & ~ & ~ & ~ & ~ & 0.6 & ~ & ~ & ~ & ~ & ~ & ~ & ~ & ~ & ~ & ~ & ~ & ~ & ~ & ~ & ~ & ~ & ~ & ~ & ~ \\ \hline
        $\text{CX}_{q_{7}; q_n}$ error (\%) & ~ & ~ & ~ & ~ & 1.3 & ~ & 0.6 & ~ & ~ & ~ & 0.5 & ~ & ~ & ~ & ~ & ~ & ~ & ~ & ~ & ~ & ~ & ~ & ~ & ~ & ~ & ~ & ~ \\ \hline
        $\text{CX}_{q_{8}; q_n}$ error (\%) & ~ & ~ & ~ & ~ & ~ & 100.0 & ~ & ~ & ~ & 0.8 & ~ & 0.5 & ~ & ~ & ~ & ~ & ~ & ~ & ~ & ~ & ~ & ~ & ~ & ~ & ~ & ~ & ~ \\ \hline
        $\text{CX}_{q_{9}; q_n}$ error (\%) & ~ & ~ & ~ & ~ & ~ & ~ & ~ & ~ & 0.8 & ~ & ~ & ~ & ~ & ~ & ~ & ~ & ~ & ~ & ~ & ~ & ~ & ~ & ~ & ~ & ~ & ~ & ~ \\ \hline
        $\text{CX}_{q_{10}; q_n}$ error (\%) & ~ & ~ & ~ & ~ & ~ & ~ & ~ & 0.5 & ~ & ~ & ~ & ~ & 0.5 & ~ & ~ & ~ & ~ & ~ & ~ & ~ & ~ & ~ & ~ & ~ & ~ & ~ & ~ \\ \hline
        $\text{CX}_{q_{11}; q_n}$ error (\%) & ~ & ~ & ~ & ~ & ~ & ~ & ~ & ~ & 0.5 & ~ & ~ & ~ & ~ & ~ & 0.8 & ~ & ~ & ~ & ~ & ~ & ~ & ~ & ~ & ~ & ~ & ~ & ~ \\ \hline
        $\text{CX}_{q_{12}; q_n}$ error (\%) & ~ & ~ & ~ & ~ & ~ & ~ & ~ & ~ & ~ & ~ & 0.5 & ~ & ~ & 3.0 & ~ & 0.7 & ~ & ~ & ~ & ~ & ~ & ~ & ~ & ~ & ~ & ~ & ~ \\ \hline
        $\text{CX}_{q_{13}; q_n}$ error (\%) & ~ & ~ & ~ & ~ & ~ & ~ & ~ & ~ & ~ & ~ & ~ & ~ & 3.0 & ~ & 1.3 & ~ & ~ & ~ & ~ & ~ & ~ & ~ & ~ & ~ & ~ & ~ & ~ \\ \hline
        $\text{CX}_{q_{14}; q_n}$ error (\%) & ~ & ~ & ~ & ~ & ~ & ~ & ~ & ~ & ~ & ~ & ~ & 0.8 & ~ & 1.3 & ~ & ~ & 1.4 & ~ & ~ & ~ & ~ & ~ & ~ & ~ & ~ & ~ & ~ \\ \hline
        $\text{CX}_{q_{15}; q_n}$ error (\%) & ~ & ~ & ~ & ~ & ~ & ~ & ~ & ~ & ~ & ~ & ~ & ~ & 0.7 & ~ & ~ & ~ & ~ & ~ & 1.4 & ~ & ~ & ~ & ~ & ~ & ~ & ~ & ~ \\ \hline
        $\text{CX}_{q_{16}; q_n}$ error (\%) & ~ & ~ & ~ & ~ & ~ & ~ & ~ & ~ & ~ & ~ & ~ & ~ & ~ & ~ & 1.4 & ~ & ~ & ~ & ~ & 0.9 & ~ & ~ & ~ & ~ & ~ & ~ & ~ \\ \hline
        $\text{CX}_{q_{17}; q_n}$ error (\%) & ~ & ~ & ~ & ~ & ~ & ~ & ~ & ~ & ~ & ~ & ~ & ~ & ~ & ~ & ~ & ~ & ~ & ~ & 1.2 & ~ & ~ & ~ & ~ & ~ & ~ & ~ & ~ \\ \hline
        $\text{CX}_{q_{18}; q_n}$ error (\%) & ~ & ~ & ~ & ~ & ~ & ~ & ~ & ~ & ~ & ~ & ~ & ~ & ~ & ~ & ~ & 1.4 & ~ & 1.2 & ~ & ~ & ~ & 0.5 & ~ & ~ & ~ & ~ & ~ \\ \hline
        $\text{CX}_{q_{19}; q_n}$ error (\%) & ~ & ~ & ~ & ~ & ~ & ~ & ~ & ~ & ~ & ~ & ~ & ~ & ~ & ~ & ~ & ~ & 0.9 & ~ & ~ & ~ & 100.0 & ~ & 1.1 & ~ & ~ & ~ & ~ \\ \hline
        $\text{CX}_{q_{20}; q_n}$ error (\%) & ~ & ~ & ~ & ~ & ~ & ~ & ~ & ~ & ~ & ~ & ~ & ~ & ~ & ~ & ~ & ~ & ~ & ~ & ~ & 100.0 & ~ & ~ & ~ & ~ & ~ & ~ & ~ \\ \hline
        $\text{CX}_{q_{21}; q_n}$ error (\%) & ~ & ~ & ~ & ~ & ~ & ~ & ~ & ~ & ~ & ~ & ~ & ~ & ~ & ~ & ~ & ~ & ~ & ~ & 0.5 & ~ & ~ & ~ & ~ & 0.9 & ~ & ~ & ~ \\ \hline
        $\text{CX}_{q_{22}; q_n}$ error (\%) & ~ & ~ & ~ & ~ & ~ & ~ & ~ & ~ & ~ & ~ & ~ & ~ & ~ & ~ & ~ & ~ & ~ & ~ & ~ & 1.1 & ~ & ~ & ~ & ~ & ~ & 0.6 & ~ \\ \hline
        $\text{CX}_{q_{23}; q_n}$ error (\%) & ~ & ~ & ~ & ~ & ~ & ~ & ~ & ~ & ~ & ~ & ~ & ~ & ~ & ~ & ~ & ~ & ~ & ~ & ~ & ~ & ~ & 0.9 & ~ & ~ & 0.7 & ~ & ~ \\ \hline
        $\text{CX}_{q_{24}; q_n}$ error (\%) & ~ & ~ & ~ & ~ & ~ & ~ & ~ & ~ & ~ & ~ & ~ & ~ & ~ & ~ & ~ & ~ & ~ & ~ & ~ & ~ & ~ & ~ & ~ & 0.7 & ~ & 1.7 & ~ \\ \hline
        $\text{CX}_{q_{25}; q_n}$ error (\%) & ~ & ~ & ~ & ~ & ~ & ~ & ~ & ~ & ~ & ~ & ~ & ~ & ~ & ~ & ~ & ~ & ~ & ~ & ~ & ~ & ~ & ~ & 0.6 & ~ & 1.7 & ~ & 0.7 \\ \hline
        $\text{CX}_{q_{26}; q_n}$ error (\%) & ~ & ~ & ~ & ~ & ~ & ~ & ~ & ~ & ~ & ~ & ~ & ~ & ~ & ~ & ~ & ~ & ~ & ~ & ~ & ~ & ~ & ~ & ~ & ~ & ~ & 0.7 & ~ \\ \hline
    \end{tabular}
    }
    }
\end{table}

\end{document}